\documentclass[12pt,a4paper]{article}
\usepackage[vcentermath,enableskew]{youngtab}
\def\slash{{\kern 2pt\raise 1.5pt \hbox{$\backslash$} \kern -9pt}}
\def\e{\hbox{e}}

\begin{document}
\title{\hfill\raise 20pt\hbox{\small DIAS-STP-03-02}\\
The Spectrum of the Dirac Operator on Coset Spaces
with Homogeneous Gauge Fields}
\author{Brian P. Dolan\\
Dept. of Mathematical Physics, NUI, Maynooth, Ireland\\
{\rm and}\\
School of Theoretical Physics, \\Dublin Institute for Advanced Studies,\\ 
10~Burlington Rd., Dublin 8, Ireland\\
{\tt bdolan@thphys.may.ie}\\}

\maketitle

\begin{abstract}
The spectrum and degeneracies of the Dirac operator are analysed
on compact coset spaces when there is a non-zero homogeneous 
background gauge field
which is compatible with the symmetries of the space, in particular
when the gauge field is derived from the spin-connection.
It is shown how the degeneracy of the lowest Landau level
in the recently proposed higher dimensional quantum Hall effect
is related to the Atiyah-Singer index theorem for
the Dirac operator on a compact coset space.
\end{abstract}

\section{Introduction}

There has long been a fruitful interplay between
condensed matter physics and field theory in
particle physics, many concepts that were first
developed in the former later being applied
to the latter and vice versa.  The quantum Hall effect \cite{Girvin}
has attracted the interest of many high energy
theorists, not least because the fractional QHE
exhibits collective excitations which mimic a fractional
electric charge, but also because there are deeper connections between
the Hall effect and string theory \cite{Bernevig}.
Recently Zhang and Hu proposed a higher
dimensional analogue of the quantum Hall effect, on $S^4$ \cite{ZhangHu},
based on Haldane's description of the Hall effect on
$S^2$ with a magnetic monopole at the centre,
\cite{Haldane}.  Zhang and Hu's idea was developed further in
\cite{HouPeng} and extended 
to complex projective spaces in \cite{KarabaliNair}.
The connection between the higher dimensional quantum Hall effect
and string theory was analysed in \cite{Fabinger}.

The higher dimensional quantum Hall effect
involves a generalisation of the Landau problem to particles
moving on a compact coset space $G/H$,
in the presence of a background gauge field: 
such as a $U(1)$ monopole on $CP^n$ or a homogeneous 
$SU(2)$ instanton field on $S^4$.  A common ingredient of the
these analyses is the calculation of the degeneracy of the ground state 
for particles moving
in a homogeneous background, i.e.~a background field which
has the symmetry of the isometry group $G$.  In previous
works, and in  this paper also, the gauge group will be
restricted to be the holonomy group $H$ (or a factor group
of same if $H$ is a product of smaller groups).

In \cite{ZhangHu,KarabaliNair} the degeneracy of the ground state
was calculated using group theory: the non-relativistic Hamiltonian for 
a spinless
particle moving in a homogeneous background field
involves the quadratic Casimirs of the groups $G$ and $H$
and the allowed states involve irreducible representations of $G$
that contain pre-ordained representations of $H$. The dimension
of the representation of $G$ corresponding to the ground state
is identified with the degeneracy of the ground state. 

It is common in discussions
of the quantum Hall effect to ignore the electron's spin.
Zhang and Hu treated scalar particles satisfying the exclusion
principle, as did Karabali and Nair:
this is perfectly justified when the Zeeman splitting 
is large enough that transitions between spin
states can be ignored and the ground state is effectively
isolated from the next highest spin state.  
Nevertheless one is tempted
to ask what is the r\^ole of electron spin in the
higher dimensional quantum Hall effect, and it will
be argued here that there is an important quantitative relic
of the Fermionic nature of the particles in the degeneracy
of the ground state, over and above the trivial
consequences of the exclusion principle.
It is shown in section 2 that
the degeneracies calculated in \cite{ZhangHu,Haldane,KarabaliNair},
for $S^4$, $S^2$ and $CP^n$ respectively, are related to the
index of the Dirac operator for Fermions moving in the appropriate
background field: in fact the degeneracy is the number of zero-modes 
of the Dirac operator and, generically, this is the modulus
of the index.  Furthermore the ground state wave-functions, the
higher dimensional analogues of the Laughlin wave-functions,
are precisely the zero-modes of the Dirac operator.

We do not have to look far to discover the reason for this --- the
square of the Dirac operator
is nothing other than the Hamiltonian 
for a non-relativistic Fermion moving in a static background field,
\begin{equation}
(i\slash D )^2=-D_\alpha D^\alpha +{R\over 4}{\bf 1} 
-{i\over 2} F_{\alpha\beta}\gamma^{\alpha\beta},
\label{NRH}
\end{equation}
where $R$ is the Ricci scalar and
the last term represents the Zeeman splitting.
(There is an extra term on the right hand side of (\ref{NRH})
if the spin connection involves torsion, this equation must therefore
be modified for
non-symmetric coset spaces with torsion as considered in section 3.)
Thus a non-relativistic particle moving in a static
background magnetic field is
an example of supersymmetric quantum mechanics with a self-dual
pre-potential \cite{susyqm}.
Since the Dirac operator is Hermitian, the eigenvalues of the
Hamiltonian (\ref{NRH}) are positive semi-definite and a zero-mode
requires exact cancellation of all three terms on the right hand side.
It is shown in sections 2 and 3 
that, for specific homogeneous background fields
(analogues of monopole and instanton fields on $S^2$ and $S^4$)
all three terms on the right hand side of (\ref{NRH}) are
mutually commuting and so can be simultaneously diagonalised
so that all spin components decouple from each other.
The Dirac Laplacian $\Delta=-D_\alpha D^\alpha$ 
is itself a positive operator on a compact space with positive curvature, 
so a zero-mode
of the Dirac operator, if one exists,
requires a cancellation of the lowest eigenvalue
of the Laplacian with the lowest eigenvalue of the sum of
curvature and Zeeman terms on the
right hand side of (\ref{NRH}).
In fact, for the homogeneous background fields that are considered
here, the eigenvalues of (\ref{NRH}) can be determined purely in terms
of certain quadratic Casimirs of the isometry group $G$ and the 
holonomy group $H$ and are given by equation (\ref{Eigenvalues}) or, 
more generally, (\ref{TorsionEigenvalues}).

Even without calculating the full spectrum 
it is possible to find the representation of $G$ with
the lowest eigenvalue
for any Fermion in a given representation of the gauge group.
If the chosen representation of the gauge group allows 
for a zero-mode of the Dirac operator then
the dimension of the representation corresponding to the
lowest eigenvalue generically gives the 
number of zero-modes.  Since this
calculation involves only the lowest eigenvalue of the Zeeman and the
curvature terms, which are fixed in advance, the Fermionic
nature of the particles can be ignored and the problem reduces
to choosing the correct representations of $G$ to scan in
minimising the Laplacian. This is precisely what was done
in \cite{ZhangHu} and \cite{KarabaliNair}.

The net result is that the number of zero-modes of the Dirac
operator, for Fermions in a given representation of the
gauge group, can be calculated simply from a knowledge of the
quadratic Casimirs of $G$ and the decompositions of its
representations under $H\mapsto G$.
This generalises the results of \cite{ZhangHu} and \cite{KarabaliNair}
to the quantum hall effect on any coset space $G/H$ with compact
Lie groups $G$ and $H$.

Of course an analysis of the spectrum of the Dirac operator
is of intrinsic interest, even without reference to the
quantum Hall effect.  In particular it is of obvious importance
in Kaluza-Klein theories and and string theory. 

The layout of the paper is as follows.  In section 2 the case
of symmetric spaces $G/H$ is treated in detail and it is shown
how the eigenvalues
of the Dirac operator in the presence of a homogeneous
background gauge field can be expressed in terms of quadratic
Casimirs $C_2(G)$ and $C_2(H)$.  The examples
of $S^2$, $S^4$ and $CP^2$ are worked out and compared to
known results.  Section 3 extends the analysis to non-symmetric
spaces with torsion and the example of $SU(3)/U(1)\times U(1)$
is treated in detail --- this space is of interest in string
theory where it is arises in the context of seven dimensional spaces
with $G_2$ holonomy and their conical singularities \cite{WA}.
Section 4 gives a summary of the results.
Some technical details are relegated to three appendices:
appendix A reviews aspects of the geometry of homogeneous spaces
used in the text.  Appendix B presents the
spectrum of the Dirac operator on $CP^2$, in the presence
of a homogeneous background $SU(2)\times U(1)$ gauge field.
Appendix C presents a
standard analysis of the Atiyah-Singer index theorem on
$SU(3)/U(1)\times U(1)$ for comparison with the
results of section 3.

\section{Symmetric Spaces}

Consider the Dirac operator for a Fermion moving on a 
$d$-dimensional compact space, without
boundary, in the presence of a background gauge field:
\begin{equation}
i\slash D = i\gamma^\alpha D_\alpha=i{e^m}_\alpha \gamma^\alpha 
\left(\partial_m + {1\over 4}\omega_{m,\alpha\beta}\gamma^{\alpha\beta}
+iA^i_m t_i\right),
\end{equation}
where $\omega_{\alpha\beta}=\omega_{\alpha\beta,m}dx^m$ is the spin connection,
$A^i=A^i_mdx^m$ the gauge connection and $t_i$ 
are generators of the gauge group ($\alpha,\beta=1,2,\cdots,d$ are
orthonormal indices and $m=1,2,\cdots,d$ is a co-ordinate index). 
The $\gamma$-matrices satisfy the usual
Clifford algebra,
\begin{equation}
\{\gamma^\alpha,\gamma^\beta\}=2\delta^{\alpha\beta},
\qquad\hbox{with}\qquad\gamma^{\alpha\beta}:={1\over 2}[\gamma^\alpha,\gamma^\beta],
\end{equation}
and ${e^m}_\alpha$ are $d$-beins for the metric.
The curvature and field strength follow from
\begin{equation}
[D_\alpha,D_\beta]=iF^i_{\alpha\beta}t_i+
{1\over 4}R_{\alpha\beta\gamma\delta}\gamma^{\gamma\delta}.
\label{FR}
\end{equation}
For a torsion free connection squaring the Dirac operator gives
\begin{equation}
(i\slash D )^2=\Delta +{R\over 4}{\bf 1} 
-{i\over 2} F_{\alpha\beta}\gamma^{\alpha\beta},
\label{Hamiltonian}
\end{equation}
with $\Delta=-D^\alpha D_\alpha$ the Dirac Laplacian.
We shall refer to $\Delta +{R\over 4}{\bf 1}$ as the kinetic
 energy
and $-{i\over 2} F_{\alpha\beta}\gamma^{\alpha\beta}=
-{i\over 2} F_{\alpha\beta}^it_i\gamma^{\alpha\beta}$ as the Zeeman
energy.
The Laplacian here is the Laplacian acting on spinors, including the spin and
the gauge connection, so its spectrum depends on both the metric and 
the background field.  

On a coset space $G/H$, with $G$ and $H$ compact groups,
it is natural to use the $G$-invariant metric, for which the generators of $G$
are Killing vectors and the holonomy group is $H\subseteq SO(d)$.  
Furthermore we shall consider background gauge fields which are compatible with 
the isometries, in the sense that Lie transport of the field strength $F$ by
a Killing vector $K$ generates a gauge transformation,
\begin{equation}
{\cal L}_K F = g^{-1}Fg,
\end{equation}
where $g\in {\cal G}$, the group of gauge transformations. 
In particular this will be the case if we identify the gauge group with the
holonomy group and the gauge connection with the spin connection --- the
details of this identification are given in appendix A.  (A variation on
this is if the holonomy group factorises into simple groups and $U(1)$ factors.
When this is the case the gauge group can be taken to be one of the factors.
For example this is the
situation for the homogeneous $SU(2)$ instanton on $S^4$, where $S^4=SO(5)/SO(4)$
and, at the level of the algebras, $H=SU(2)\times SU(2)$ so we can take
the gauge group to be just $SU(2)$.)

Let $t_A$ be the generators of the isometry group $G$, with $[t_A,t_B]=i{f_{AB}}^C t_C$,
and $t_i$ the generators of the holonomy group $H$.
Then the curvature 2-forms of the $G$-invariant metric for a symmetric 
space are (see appendix A),
\begin{equation}
{R^\alpha}_\beta = {1\over 2}{R^\alpha}_{\beta\gamma\delta}\e^\gamma\wedge\e^\delta
={1\over 2} {f^\alpha}_{\beta i}{f^i}_{\gamma\delta}\e^\gamma\wedge\e^\delta.
\end{equation}
Identifying the gauge connection with the spin connection gives rise to the
field strength,
\begin{equation}
F^i={1\over 2}F^i_{\alpha\beta}\e^\alpha\wedge\e^\beta=
{1\over 2}{f^i}_{\alpha\beta}\e^\alpha\wedge\e^\beta.
\end{equation}
For a symmetric space the Riemann tensor is co-variantly 
constant and this means that the
above field strength is co-variantly constant,
\begin{equation}
D_\alpha F^i_{\beta\gamma}=0.
\end{equation}
In particular the Laplacian commutes with the Zeeman term in the Hamiltonian.

With this choice of background field the commutator (\ref{FR}) simplifies,
\begin{equation}
[D_\alpha,D_\beta]=i{f^i}_{\alpha\beta}\left\{\bigl({\bf 1}\otimes t_i\bigr) - 
{i\over 4}f_{i\gamma\delta}\bigl(\gamma^{\gamma\delta}\otimes{\bf 1}\bigr)\right\}.
\end{equation}
Now 
\begin{equation}
T_i:=-{i\over 4}f_{i\gamma\delta}\gamma^{\gamma\delta}
\end{equation}
are a representation
of the gauge group (which may be reducible, in general),
\begin{equation}
[T_i,T_j]=i{f_{ij}}^k T_k,
\end{equation}
so 
\begin{equation}
[D_\alpha,D_\beta]={f^i}_{\alpha\beta} D_i
\end{equation}
with
\begin{equation}
D_i:=i\bigl\{( {\bf 1}\otimes t_i) + ( T_i\otimes {\bf 1})\bigr\}
\label{Di}
\end{equation}
being the generators of $H$ in the tensor product representation of
$t_i$ with the spinor representation $T_i$.
This allows the Laplacian to be expressed as the difference of quadratic Casimirs,
\begin{equation}
\Delta=-D_\alpha D_\alpha = -D_A D_A + D_i D_i =C_2(G,\cdot)-C_2(H,D_i).
\label{Laplacian}
\end{equation}
For spinors in a given representation $t_i$ of the gauge group $C_2(H,D_i)$
in this expression is always calculated in the fixed
representation (\ref{Di}),
which in general involves reducible representations of $G$,
while the representations used in $C_2(G,\cdot)$ 
range over all irreducible representations
of $G$ than contain (\ref{Di}).
In particular the cross-term $-2t_i\otimes T_i$ from $(D_i)^2$ 
in (\ref{Laplacian}) exactly cancels
the Zeeman energy in (\ref{Hamiltonian}) and, as described in
appendix A, the second order Casimir for the representation $T_i$
is related to the Ricci scalar by $C_2(H,T_i)=R/8$.
The eigenvalues of the square of the Dirac operator (\ref{Hamiltonian}) 
can then be expressed purely in terms of quadratic Casimirs:
\begin{equation}
E=C_2(G,\cdot)-C_2(H,t_i)+{R\over 8}{\bf 1}.
\label{Eigenvalues}
\end{equation}

This construction will now be illustrated with some examples.
\bigskip

\noindent {\it i) $S^2\cong SO(3)/SO(2)$}

\noindent This was the geometry originally studied by Haldane in the
context of the quantum Hall effect \cite{Haldane}.
The isometry group is generated by the algebra of $SU(2)$
\begin{equation} [t_A,t_B]=i{\epsilon_{AB}}^C t_C
\end{equation}
and we are free to choose $t_3$ to generate the $U(1)$
holonomy. Formula (\ref{curvature}) of appendix A gives
\begin{equation}
R_{\alpha\beta}
={1\over 2}\epsilon_{\alpha\beta3}\epsilon^{3\gamma\delta}\e^{\gamma}\wedge\e^{\delta}
\end{equation}
so 
\begin{equation}
R_{12}=\e^1\wedge\e^2
\end{equation}
are the curvature 2-forms for a sphere of unit radius.
Also
\begin{equation}
F^3=\e^1\wedge\e^2
\end{equation}
is the field strength if a magnetic monopole at the centre of the sphere.
Actually this corresponds to a monopole of charge 2, since
\begin{equation}
{1\over 2\pi}\int_{S^2}\e^1\wedge\e^2=2
\end{equation}
is the Chern class of the tangent
bundle (which is equal to the Euler characteristic).
In general we can put a monopole of any integral charge at the centre
of the sphere
\begin{equation}
F^3={M\over 2}\e^1\wedge\e^2.
\end{equation}
(Alternatively we can work with a monopole of charge 2 and consider
Fermions of any half-integral charge in this background.) 

Choosing $\gamma^1=\sigma^1$ and $\gamma^2=\sigma^2$, with $\sigma^1$ and
$\sigma^2$ Pauli matrices,  we have
\begin{equation}
{i\over 2}F^3_{\alpha\beta}\gamma^{\alpha\beta}=iF^3_{12}(i\sigma^3)
=-{M\over 2}\left(\matrix{1&0\cr 0&-1}\right).
\end{equation}
The Ricci scalar for a sphere of unit radius is 2, so equation 
(\ref{Hamiltonian}) gives
\begin{equation}
(i\slash D)^2=\Delta +{1\over 2}{\bf 1}+{M\over 2}\left(\matrix{1&0\cr 0&-1}\right).
\label{S2Hamiltonian}
\end{equation}
For positive $M$ this
indicates that there are spin down zero-modes of the Dirac operator
if
\begin{equation}
\Delta+{1\over 2}={M\over 2}
\end{equation}
while for negative $M$ there are spin up zero-modes
if
\begin{equation}
\Delta+{1\over 2}=-{M\over 2}.
\end{equation}
There are of course no zero-modes for $M=0$ as required by
Lichnerowicz theorem.

In this example $t_i$ of (\ref{Di}) is just a number, $M/2$,
and $T_i$ is $\sigma_3/2$ 
so 
\begin{equation}
D_3=i\left(\matrix{ {M+1\over 2} & 0\cr 0& {M-1\over 2}\cr}\right)
\qquad\Rightarrow\qquad D_3D_3=-
\left(\matrix{ \left({M+ 1\over 2}\right)^2&0\cr 0& \left({M- 1\over 2}\right)^2\cr}\right).
\end{equation}
The eigenvalues of the Laplacian (\ref{Laplacian}) are therefore
\begin{equation}
\Delta_j=j(j+1)-\left({M\pm 1\over 2}\right)^2,
\end{equation}
as discussed in \cite{ZhangHu},
so eigenvalues of (\ref{S2Hamiltonian}) are 
\begin{equation}
E_j={(2j+1)^2-M^2\over 4},
\end{equation}
which can also be obtained directly from (\ref{Eigenvalues}).
For $M=0$ this reproduces the well-known result that the
spectrum of the Dirac operator is linear in angular momentum
(see e.g. \cite{BGJP}).

For $M\ne 0$ the representations $j$ of $SU(2)$ that appear in a harmonic
expansion of $\Delta$ are restricted to those that contain the 
$U(1)$ representation of charge ${M\pm 1\over 2}$,
i.e.~$j={\left|M\right|-1\over 2} +k$ with $k$ a non-negative integer,
\begin{equation}
E_j=k(k+|M|).
\end{equation}
There are zero-modes for $k=0$,
and $j_{min}={M-1\over 2}$ for positive $M$ or
$-({M+1\over 2})$ for negative $M$.
In either case the degeneracy of the ground state is
\begin{equation}
d(j_{min})=2j_{min}+1=|M|,
\label{degeneracy}
\end{equation}
which is the number of zero-modes of the
Dirac operator.  Note the shift of $j_{min}$ away from $|M|$
by $1/2$, due to the intrinsic spin of the Fermion.

The degeneracy (\ref{degeneracy}) relates to the Atiyah-Singer index theorem which states that
the the index of the Dirac operator is minus the first Chern class
\cite{EGH},
\begin{equation}
\nu=\nu_+-\nu_-=-{1\over 2\pi}\int_{S^2}F^3=-M,
\end{equation}
where $\nu_+$ is the number of positive chirality zero-modes
and $\nu_-$ the number
of negative chirality zero-modes.
Indeed the ground state wave-functions in \cite{Haldane}
for the integer quantum Hall effect, spherical analogues of the
Laughlin wave-functions, are precisely these zero-modes.
The case $|M|=1$  corresponds to $j_{min}=0$,
in this case the gauge connection exactly cancels the spin connection
for the relevant chirality and single zero-mode
of the Dirac operator is a constant spinor.

The above calculation can be represented graphically
using Young tableaux, which will be useful in more complicated situations
to follow.
The fundamental of $SU(2)$ decomposes as
\begin{equation}
SU(2)\rightarrow U(1)\qquad {\bf 2}\rightarrow {\bf 1}_1+{\bf 1}_{-1}.
\end{equation}
Denoting ${\bf 1}_1$ by $\vcenter{\hbox{\tiny$\young(\times)$}}$
and ${\bf 1}_{-1}$ by $\vcenter{\hbox{\tiny$\young(\bullet)$}}$ 
the $(p+1)$-dimensional irreducible representation of $SU(2)$ contains
\def\Dots{\cdot\cdot}
\begin{equation}
\underbrace{\young(\times \Dots \times )}_{s}\kern -2.5pt
\underbrace{\young(\bullet \Dots \bullet )}_{r}
\subset\underbrace{\young(\ \Dots \ )}_{p}
\end{equation}
with $p=r+s$.  Fixing the $U(1)$ charge to be $Q$ constrains $s-r=Q$ 
so $p=2r+Q$. The ground state energy for a Fermion in this
background can now be
found by minimising $\Delta$, since all the other terms in the
energy are constants for fixed $Q$, that is by minimising
\begin{equation}
{p\over 2}\left({p\over 2}+1\right)=r(r+1)+{Q\over 2}(2r+1) + {Q^2\over 4}.
\end{equation}
If $Q>0$ this is minimised by $r=0$, so $p=Q$ and the
degeneracy of the ground state is $Q+1$.
If $Q<0$ it is minimised by  $r=p$ and then $p=-Q$, so the
degeneracy is $-Q+1$.  In either case the ground state has 
$p=|Q|$ and the degeneracy
is $|Q|+1$.  Clearly $|Q|=2j_{min}$ and the $U(1)$ charge $Q$
is not just the monopole charge $M$, but includes a shift to
account for the intrinsic spin of the Fermion, $|Q|=|M|-1$.

This method, using representation theory to describe the kinetic
energy and calculate the degeneracy of the
ground state, was used in \cite{KarabaliNair}:
though in that reference the particles were treated as scalars
so there was no intrinsic spin --- there was therefore no Zeeman
energy to make the total ground state energy vanish and no shift in the
charge to account for the intrinsic spin of the particles.
The technique is however applicable to both the 
Laplacian for scalars and the
square of the Dirac operator because, for a given gauge background
and representation,
they only differ by constants.  It has the advantage of avoiding 
an explicit calculation
of the full eigenvalue spectrum of the Dirac operator.

\bigskip
\noindent {\it ii) $S^4\cong SO(5)/SO(4)$}

\noindent The next example, $S^4$, was the case studied in the
first paper on the higher dimensional quantum Hall effect, \cite{ZhangHu}.
In this case the algebra of the holonomy group is $SU(2)\times SU(2)$
and we can take the gauge group to be just one $SU(2)$ factor.
The Riemann tensor can be split into self-dual and anti-self-dual
parts and these correspond to the curvatures arising from the
two $SU(2)$ factors of the holonomy group.  Choosing, for example, the
self-dual $SU(2)$ factor the resulting $
SU(2)$ background gauge field
is the homogeneous instanton of charge one, which has
$SO(5)$ symmetry on $S^4$ \cite{Yang} (this paper
was published a little after the BPST instanton \cite{BPST}, 
but the techniques
are very enlightening and highlight the analogy with the
Wu-Yang monopole -- Yang calls this homogeneous instanton
configuration a non-abelian monopole).

Representations of $SO(5)$ can be labelled by two integers
$p$ and $q$ with $p\ge q$.  The second order Casimir and dimension
are given by
\begin{equation}
C_2(p,q)={p^2 +q^2\over 2}+2p + q
\end{equation}
and
\begin{equation}
d(p,q)={1\over 6}(p+q+3)(p-q+1)(p+2)(q+1)
\end{equation}
respectively.

Now suppose we have a particle on $S^4$ in the representation $I$ of
$SU(2)$ in the background of a homogeneous instanton.
Demanding that an $SO(5)$ irreducible representation contains
the $I$ of $SU(2)$ implies \cite{Yang}
\begin{equation}
p-q=2I,
\end{equation}
and so
\begin{equation}
C_2(q+2I,q)=q^2+q(2I+3)+2I^2+4I.
\end{equation}
The Ricci scalar for the unit four-sphere is $R=12$ so
the eigenvalues (\ref{Eigenvalues}) of $(i\slash D)^2$ for a Fermion in the representation
$J$ of the gauge group are thus

\begin{equation}
E=q^2+q(2I+3)+2I^2+4I-2J(J+1)+{3\over 2}.
\end{equation}
(The factor of two in front of the gauge Casimir $J(J+1)$ here is due
to the fact that the Dirac operator is non-chiral,
the holonomy group is $SU(2)\times SU(2)$, and both chiralities
couple to the gauge group in the same way.)
The total isospin $I$ is a combination of the gauge isospin $J$
and the intrinsic spin of the Fermion, $I=J\pm 1/2$, so the
energy levels are labelled by the integer $q$ and
\begin{eqnarray}
E_+(q)&=&q^2+q(2I+3)+2(2I+1)\nonumber\\
E_-(q)&=& q^2+q(2I+3)
\label{Spherespectrum}
\end{eqnarray}
both with degeneracies  
\begin{equation}
d(2I+q,q)={1\over 6}(2q+2I+3)(2I+1)(q+2I+2)(q+1).
\end{equation}

The spectrum of the Dirac operator on $S^4$ has already been studied in the
literature.  With $J=0$, so $I=1/2$, (\ref{Spherespectrum}) reduces to
\begin{equation}
E_+(q)=(q+2)^2.
\end{equation}
The eigenvalues of the Dirac operator itself are therefore linear
in $q$ and reproduce the result of \cite{BGJP}, with $q=j-1/2$
in their notation.

Zero-modes of the Dirac operator require $J\ne 0$ and $I=J-1/2$
so that $q=0$ has vanishing eigenvalue, $E_-(0)=0$.
The degeneracy of this ground state is
\begin{equation}
d(2I,0)={1\over 6}(2I+3)(2I+2)(2I+1)
\label{Sfourdegeneracy}
\end{equation}
and these are precisely the degeneracies found in \cite{ZhangHu}
for the ground state of the higher dimensional quantum Hall effect
on $S^4$.  This can be related to the index of the Dirac operator.
Recall that the $SU(2)$ instanton
in the $J$ representation can be represented by the field strength
\begin{equation}
F=F^i t_i\qquad\hbox{with}\qquad F^i=\e^i\wedge\e^4 +{1\over 2}\epsilon^{ijk}
\e^j\wedge\e^k
\end{equation}
and $\sum_it_it_i=J(J+1){\bf 1}$, $t_i$ being 
$(2J+1)\times(2J+1)$-dimensional matrices
(see \cite{EGH} for example).
The index of the Dirac operator for a Fermion in the $J$-representation
of the gauge group $SU(2)$ is then
\begin{equation}
\nu={1\over 2(2\pi)^2}tr\int_{S^4}F\wedge F={1\over 4\pi^2}tr(t_it_i)\int_{S^4}
\e^1\wedge\e^2\wedge\e^3\wedge\e^4 ={2\over 3}(2J+1)J(J+1),
\end{equation}
since the unit $S^4$ has volume $\int_{S^4}\e^1\wedge\e^2\wedge\e^3\wedge\e^4=
8\pi^2/3$.  This agrees with the degeneracy of the ground state 
(\ref{Sfourdegeneracy}) if
\begin{equation}
I=J-{1\over 2}.
\label{JI}
\end{equation}
The representation $I$ again includes the intrinsic spin of the
Fermion: $I$ is in the tensor product of the $J$ with the $1/2$ in
(\ref{Di}).  Thus $J=1/2$ can give $I=0$ and
the single zero-mode of the Dirac operator 
for a Fermion in the fundamental representation
of $SU(2)$ is a constant spinor on $S^4$ because the gauge connection
again exactly cancels the spin connection in this case.
The analogues on $S^4$ of the Laughlin-Haldane wave-functions on $S^2$
are constructed in \cite{ZhangHu} by taking anti-symmetrised products
of the zero-modes of the Dirac operator in the instanton background
with $I=1/2$.
\bigskip

\noindent {\it iii) $CP^2\cong SU(3)/U(2)$}

\noindent Our final example of a symmetric space is $CP^2$.  This is not
a spin manifold, there is a topological obstruction to defining
spinors on this space, but spinors can exist if they are
coupled to an appropriate background gauge field \cite{HP}.
The gauge group is taken to be $SU(2)\times U(1)$ and the representations
are labelled by the isospin $I$ and the hypercharge $Y$.
Representations of $SU(3)$ which contain the $(I,Y)$ representation
of $SU(2)\times U(1)$ are easily found. 
Under $SU(3)\rightarrow SU(2)\times U(1)$ the ${\bf 3}$ and  $\bar {\bf 3}$
decompose as 
\begin{equation}
{\bf 3}\rightarrow {\bf 2}_1 + {\bf 1}_{-2}
\qquad\hbox{and}\qquad
\bar{\bf 3}\rightarrow \bar{\bf 2}_{-1} + {\bf 1}_{2}.
\end{equation}
Let $\vcenter{\hbox{\tiny$\young(\times)$}}={\bf 2}_1$, 
$\vcenter{\hbox{\tiny$\young(\bullet)$}}={\bf 1}_{-2}$,
$\vcenter{\hbox{\tiny$\young(\times,\times)$}}={\bf 1}_2$ and 
$\vcenter{\hbox{\tiny$\young(\bullet,\times)$}}={\bf 2}_{-1}$.
A general irreducible representation of $SU(3)$ can be labelled
by two integers $p$ and $\bar p$ and it contains
\begin{equation}
\underbrace{\young(\times\Dots\times,\times\Dots\times)}_{\bar r}\kern -2.3pt
\underbrace{\young(\bullet\Dots\bullet,\times\Dots\times)}_{\bar s}\kern -2.5pt
\raise 6.7pt \hbox{$\underbrace{\young(\bullet \Dots \bullet )}_{r}\kern -2pt
\underbrace{\young(\times \Dots \times )}_{s}$}
\qquad\subset\qquad
\underbrace{\young(\ \Dots \ ,\ \Dots \ )}_{\bar p}\kern -2.2pt
\raise 6.7pt\hbox{$\underbrace{\young(\ \Dots \ )}_{p}$}
\end{equation}
with $r+s=p$, $\bar r + \bar s =\bar p$.  
The second order Casimir and dimension are
\begin{equation}
C_2(p,\bar p)={1\over 3}\Bigr(p(p+3) + \bar p(\bar p+3) + p\bar p\Bigl)
\label{SU3Casimir}
\end{equation}
and
\begin{equation}
d(p,\bar p)={1\over 2}(p+\bar p +2)(p+1)(\bar p+1)
\label{SU3dimension}
\end{equation}
respectively.
Furthermore there are constraints
\begin{equation}
(s -\bar s) -2(r-\bar r)=Y \qquad \hbox{and} \qquad{s+\bar s\over 2}=I
\end{equation}
 for $(p,\bar p)$
to contain $(I,Y)$.  
This means that 
the allowed states are labelled by two integers, as first noted in 
\cite{KarabaliNair}.  Note that $Y$ is odd if $I$ is half-integral and
even if $I$ is integral.

The full spectrum of the square of the Dirac operator is given
in appendix B, here we shall concentrate on the zero-modes.
There are two cases to consider: $|Y|\ge 2I$ and  $|Y|\le 2I$:
\begin{itemize}
\item For $|Y|> 2I$ equations (\ref{Yge2I}) and (\ref{spindown}) 
indicate that there are zero-modes with degeneracy
  \begin{equation}
   d(0,0)={1\over 8}(2I+1)(|Y|-2I+2)(|Y|+2I+4)
\label{degeneracya}
  \end{equation}
\item For $|Y|< 2I$ equations (\ref{Yle2I}) and (\ref{spinup4})
indicate that there are zero-modes with degeneracy
  \begin{equation}
   d(0,0)={1\over 4}(I+1)(2I+2+Y)(2I+2-Y)
\label{degeneracyb}
  \end{equation}
\end{itemize}
For $Y=0$ equation (\ref{degeneracyb}) agrees with the degeneracy of the
lowest Landau level for a scalar field derived in \cite{KarabaliNair}.
Although (\ref{degeneracya}) differs from the corresponding expression
for scalars in \cite{KarabaliNair}, it agrees for $|Y|\rightarrow\infty$,
which is the limit used in their analysis.

But now
equation (\ref{degeneracya}) can be compared with known results for the Dirac index
on $CP^2$.
For $I=0$ we have a $U(1)$ bundle with $Y$ even and
\begin{equation}
d(0,0)={(|Y|+2)(|Y|+4)\over 8},
\end{equation}
while the index is  \cite{HP}
\begin{equation}
\nu={(Q+1)(Q+2)\over 2},
\end{equation}
with $Q$ the $U(1)$ charge of the Fermion, including a spin
contribution.
So $Q>0$ requires $Y=2Q$, while $Q\le -3$ requires $Y=2Q+6$.
$Q=0$ implies $Y=0$ and
is the so-called spin$^c$ structure, where the spin exactly cancels the
monopole charge.  For the particular cases $Q=-1$ and $Q=-2$ there are
generically no zero-modes of the Dirac operator.

For $I=1/2$ we have a rank-2 vector bundle with $Y$ odd and
\begin{equation}
d(0,0)={(|Y|+1)(|Y|+5)\over 4},
\end{equation}
while the index is \cite{DolanNash}
\begin{equation}
\nu=(Q+1)(Q+3).
\end{equation}
So $Y=2Q+1$ when $Q\ge 0$ and $Y=2Q+7$ when $Q<-4$.  For $Q=-1$ or $Q=-3$
there are again generically no zero-modes.  $Q=-2$ has $\nu=-1$
and so degeneracy $d(0,0)=1$; it is therefore a singlet of $SU(2)$ 
and requires $I=0$ and $Y=0$: the Fermion spin cancels against the isospin
and monopole charge
(analogous to $J=1/2$ in (\ref{JI})).

\bigskip

\section{Non-symmetric Spaces}

On a non-symmetric coset space the analysis of section 2 requires some
minor modifications.  On non-symmetric spaces
there is a natural spin connection  which has torsion, as
described in appendix A, and equation (\ref{Hamiltonian}) is 
modified to
\begin{equation}
(i\slash D )^2=\Delta +{R\over 4}{\bf 1} -
{1\over 8}R_{\alpha\beta\gamma\delta}\gamma^{\alpha\beta\gamma\delta}
-{i\over 2} F_{\alpha\beta}\gamma^{\alpha\beta},
\label{TorsionHamiltonian}
\end{equation}
because ${R^\alpha}_{[\beta\gamma\delta]}\ne 0$ when there is torsion.
Indeed for non-symmetric coset spaces there is a spin connection with
\begin{equation}
{R^\alpha}_\beta\wedge\e^\beta=dT^\alpha+{\omega^\alpha}_\beta\wedge T^\beta,
\end{equation}
where
\begin{equation}
T^\alpha = {1\over 2}{f^\alpha}_{\beta\gamma}\e^\beta\wedge\e^\gamma
\end{equation}
are the torsion 2-forms.  The connection involving torsion is
used here because the resulting curvature tensor 
\begin{equation}
{R^\alpha}_\beta={1\over 2}{f^\alpha}_{\beta i}{f^i}_{\gamma\delta}
\e^\gamma\wedge\e^\delta
\end{equation}
is co-variantly
constant with this connection, as shown in appendix A.  This means
that at least the first three terms on the right hand side of equation 
(\ref{TorsionHamiltonian}) are mutually commuting and simultaneously
diagonalisable.

Identifying the gauge connection with the spin connection now gives the
same expression as for symmetric spaces
\begin{equation}
F^i={1\over 2}{f^i}_{\alpha\beta}\e^\alpha\wedge\e^\beta
\end{equation}
and we see that
\begin{equation}
R_{\alpha\beta\gamma\delta}\gamma^{\alpha\beta\gamma\delta}
=f^i_{\alpha\beta}f^i_{\gamma\delta}\gamma^{\alpha\beta\delta\gamma}
\end{equation}
is related to $tr(F\wedge F)$.
Furthermore
\begin{equation}
f^i_{\alpha\beta}f^i_{\gamma\delta}\gamma^{\alpha\beta\delta\gamma}
=f^i_{\alpha\beta}f^i_{\gamma\delta}\gamma^{\alpha\beta}\gamma^{\gamma\delta}
+2f^i_{\alpha\beta}f^i_{\alpha\beta}
=-16C_2(H,T_i)+2R
\end{equation}
commutes with $F^i_{\alpha\beta }\gamma^{\alpha\beta }$, ${\bf 1}$ and $\Delta$ (the last because $F^i_{\alpha\beta}$
is co-variantly constant).
Now every term on the right hand side of (\ref{TorsionHamiltonian})
is mutually commuting and it reads
\begin{equation}
(i\slash D )^2=\Delta  + 2C_2(H,T_i)
-{i\over 2} F_{\alpha\beta}\gamma^{\alpha\beta}.
\label{TorsionHamiltonian2}
\end{equation}

Paralleling the argument that led to (\ref{Eigenvalues})
for symmetric spaces the eigenvalues can be expressed terms of Casimirs
\begin{equation}
E=C_2(G,\cdot)-C_2(H,t_i)+C_2(H,T_i).
\label{TorsionEigenvalues}
\end{equation}
This equation is more general than (\ref{Eigenvalues})
and includes it as a special case,
because $C_2(H,T_i)=R/8$ for symmetric coset spaces.

As an illustration consider the space $SU(3)/ U(1)\times U(1)$,
which is of interest in string theory, where a seven dimensional space
with $G_2$ holonomy can have a conical singularity on $SU(3)/ U(1)\times U(1)$,
\cite{WA}.  We shall not calculate the full spectrum of the Dirac
operator here, but shall concentrate on the zero-modes.
There are two independent $U(1)$ gauge
fields and we can take ``monopoles'' of both, so the background
gauge field is labelled by two integers $Y$ and $T$, which are the
hypercharge and the third component of isospin of a $SU(3)$ irreducible
representation.  
Under
\begin{displaymath}
SU(3) \qquad 
\rightarrow \qquad SU(2)\times U(1) \qquad\rightarrow \qquad U(1)\times U(1)
\end{displaymath}
The ${\bf 3}$ of $SU(3)$ decomposes as
\begin{displaymath}
\qquad\quad {\bf 3} \qquad\rightarrow \qquad{\bf 2}_1 + {\bf 1}_{-2} \qquad\rightarrow
\qquad (1,1)+(-1,1)+(0,-2).
\end{displaymath}
Let $\vcenter{\hbox{\tiny$\young(\times)$}}=(1,1)$, 
$\vcenter{\hbox{\tiny$\young(\ )$}}=(-1,1)$, 
$\vcenter{\hbox{\tiny$\young(\bullet)$}}=(0,-2)$,
$\vcenter{\hbox{\tiny$\young(\times,\ )$}}=(0,2)$.
Then $\vcenter{\hbox{\tiny$\young(\bullet,\times)$}}=(1,-1)$ and 
$\vcenter{\hbox{\tiny$\young(\bullet,\ )$}}=(-1,-1)$.
The $(p,\bar p$) representation of $SU(3)$ contains
\begin{equation}
\underbrace{\young(\times\Dots\times,\  \Dots \ )}_{\bar r}\kern -2.3pt
\underbrace{\young(\bullet\Dots\bullet,\times\Dots\times)}_{\bar s}\kern -2.5pt
\underbrace{\young(\bullet\Dots\bullet,\ \Dots\ )}_{\bar t}\kern -2.5pt
\raise 6.7pt \hbox{$\underbrace{\young(\bullet \Dots \bullet )}_{r}\kern -2pt
\underbrace{\young(\  \Dots \  )}_{s}\kern -2pt
\underbrace{\young(\times \Dots \times )}_{t}$}
\qquad\subset\qquad
\underbrace{\young(\ \Dots \ ,\ \Dots \ )}_{\bar p}\kern -2.2pt
\raise 6.7pt\hbox{$\underbrace{\young(\ \Dots \ )}_{p}$}
\end{equation}
with $r+s+t=p$, $\bar r + \bar s + \bar t=\bar p$.  
Demanding that the $(p,\bar p)$ representation of $SU(3)$ contains
the $(T,Y)$ representation of $U(1)\times U(1)$ puts
constraints on the six integers $(r,s,t;\bar r,\bar s,\bar t)$, namely
\begin{equation}
Y=(s-\bar s) + (t - \bar t) -2(r-\bar r)\qquad\hbox{and}\qquad
T=(t-\bar t)-(s-\bar s).
\end{equation}
The $SU(3)$ states that contain $(T,Y)$ are therefore 
labelled by four independent integers
(note that $Y$ and $T$ are either both even or both odd).
If zero-modes of the Dirac operator exist, then their number
is given by the dimension of the $SU(3)$ representation
which minimises the second order Casimir (\ref{SU3Casimir}),
subject to these constraints.  We find
\begin{eqnarray*}
T\ge Y\ge 0\Rightarrow&
C_2^{min}
=T+{T^2\over 4}+{Y^2\over 12},\ \hbox{for
${(r,s,t;\bar r,\bar s,\bar t) =(0,0,{T+Y\over 2};0,{T-Y\over 2},0)}$}\\
Y\ge T\ge 0\Rightarrow& 
C_2^{min}
={T+Y\over 2}+{T^2\over 4}+{Y^2\over 12}, \ \hbox{for
{$(r,s,t;\bar r,\bar s,\bar t) = (0,0,T;{Y-T\over 2},0,0).$}}\\
\end{eqnarray*}
These two representations have dimensions
\begin{equation}
d_{min}={1\over 8}(T+Y+2)(T-Y+2)(T+2) \quad\hbox{and}\quad
d_{min}={1\over 8}(Y+T+4)(Y-T+2)(T+1)
\end{equation}
respectively.
Other combinations of signs follow from
\begin{eqnarray*}
T&\rightarrow& -T,\ Y\rightarrow -Y \ \,\hbox{interchange} 
\quad (r,s,t)\leftrightarrow
(\bar r,\bar s,\bar t)\\
T&\rightarrow& -T,\ Y\rightarrow \ Y \quad \hbox{interchange} 
\quad (s,\bar s)\leftrightarrow
(t,\bar t)\\
T&\rightarrow& \ T,\ Y\rightarrow -Y \quad \hbox{interchange} 
\quad (r,s,t)\leftrightarrow (\bar r, \bar t,\bar s).\\
\end{eqnarray*}
In summary the dimensions of the ground states are
\begin{eqnarray*}
|T|\ge |Y| &\Rightarrow&
d_{min}={1\over 8}\Bigl((|T|+2)^2-Y^2\Bigr)(|T|+2)\\
|Y|\ge |T| &\Rightarrow& 
d_{min}={1\over 8}\Bigl((|Y|+3)^2 - (|T|+1)^2\Bigr)(|T|+1).\\
\end{eqnarray*}
That these do indeed correspond to zero-modes can be established
by comparison with a more standard
analysis (as presented in appendix C for example): for $U(1)$
fields with monopole charges $M$ (hypercharge) and $N$ 
(third component of isospin), 
the index is (\ref{su3index})
\begin{equation}
\nu={1\over 8}M(N^2-M^2).
\end{equation}
Clearly 
\begin{equation}
M=T\pm 2 \quad\hbox{and}\quad N=Y, \qquad\hbox{for}\quad |T|\ge |Y|
\end{equation}
and
\begin{equation}
M=T\pm 1 \quad\hbox{and}\quad  N=Y\pm 3, \qquad\hbox{for}\quad |Y|\ge |T|
\end{equation}
will do the trick
(either identification works when $|T|=|Y|$).
This shift is again due to the intrinsic spin of the
Fermions contributing to the $U(1)$ charges.  

\section{Conclusions}

In this paper the spectrum of the Dirac operator, for Fermions
coupled to topology non-trivial homogeneous 
background gauge fields on compact coset spaces
$G/H$, has been determined purely in terms of quadratic Casimirs.
The spectra are obtained from (\ref{TorsionEigenvalues}),
which reduces to (\ref{Eigenvalues}) for symmetric spaces.
This is the central result.

It has been shown how the ground state degeneracies 
encountered in discussions of the
quantum Hall effect on $S^2$, and its higher dimensional
generalisations to $S^4$ and $CP^n$,
are related to the Atiyah-Singer index
theorem for spinors on these spaces.
Physically the square of the
Dirac operator is the non-relativistic Hamiltonian
for a particle moving in a background ``magnetic'' field and consists
of a kinetic term and a Zeeman splitting term (at least for
zero torsion).
For spinors on a coset space $G/H$ and gauge group 
$H$,\footnote{Or a factor group if $H$ is a product of Lie groups.}
moving in a homogeneous background field identified with the natural spin
connection, the kinetic term and the Zeeman splitting term commute.
The Hamiltonian can then be diagonalised resulting in
a set of uncoupled spin states,
each of which can be treated independently.

If all we are interested in is zero-modes and the index theorem
it is not necessary to compute the whole spectrum.
If zero-modes of the Dirac operator exist the ground state
must be one in which the Zeeman energy exactly cancels the
kinetic energy and its degeneracy is the number of zero-modes.
Since the kinetic energy is positive-definite,
this cancellation requires a minimum of the quadratic Casimir corresponding
to the kinetic energy.  The degeneracy of the ground state is
then just the dimension of the representation of $G$
that minimises the quadratic Casimir --- subject to the condition
that only representations of $G$ that contain the representations
of $H$ relevant to the spinor dynamics are considered.
Generically the number of zero-modes is the same as 
the modulus of index of the
Dirac operator, so this gives a method of calculating the
index just by using knowledge of the quadratic Casimir.

The construction is not one-to-one: there may be more than
one possible representation of the gauge group giving the same
number of zero-modes of Dirac operator.  For example on
$CP^2$ an $SU(2)$ singlet with $Q=-3$ has $|\nu|=1$ which
is the same as for an $SU(2)$ doublet with $Q=-2$.
The complete spectrum of the Dirac operator on $CP^2$
is given in appendix B.

The technique also works for non-symmetric space with torsion
and the example of $SU(3)/U(1)\times U(1)$ has been treated in detail.

\smallskip
\leftline {\bf Acknowledgements}

\smallskip
It is a pleasure to thank Charles Nash for useful discussions
on the Dirac operator and the Atiyah-Singer index theorem.

\bigskip
\appendix

\section{Appendix}

In this section we summarise the construction of
homogeneous connections and curvatures on coset spaces.
None of this material is new: the construction is well known
to mathematician's
\cite{KobayashiNomizu}.
The development here is based on the exposition for physicists in
\cite{SalamStrathdee}.

On coset spaces $G/H$, with $G$ and $H$ compact Lie groups and $G$ simple,
the Riemann tensor can be obtained in terms of the
structure constants of $G$.  
Denote the generators of $G$ by $t_A$, with the algebra
\begin{equation}
[t_A,t_B]=i{f_{AB}}^Ct_C.
\end{equation}
The set of generators $\{t_A\}$ decomposes
into generators of $H$, which
shall be denoted by $t_i$, and the remaining generators, which
will be labelled $t_\alpha$.
Thus $\alpha$ takes on $d$-values where $d$ is the dimension
of $G/H$.
The algebra of $G$ now splits up as
\begin{equation}
[t_i,t_j]=i{f_{ij}}^kt_k\qquad
[t_i,t_\alpha]=i{f_{i\alpha}}^\beta t_\beta \qquad
[t_\alpha,t_\beta]=i{f_{\alpha\beta}}^\gamma t_\gamma+i {f_{\alpha\beta}}^k t_k.
\label{GHalgebra}
\end{equation}
The space $G/H$ is called symmetric if ${f_{\alpha\beta}}^\gamma =0$,
when this is the case the decomposition (\ref{GHalgebra}) is 
invariant under $t_\alpha \rightarrow -t_\alpha$.

Orthonormal 1-forms for the group $G$ can be chosen as the Cartan 1-forms,
\begin{equation}g^{-1}dg =ie^A t_A, \end{equation}
so
\begin{equation}de^A={1\over 2}{f_{BC}}^A e^B\wedge e^C.\end{equation}

Then the subset $e^\alpha$ are orthonormal 1-forms for the $G$-invariant
metric on $G/H$ and the remaining 1-forms $e^i$ can be
expanded on $G/H$ as $e^i={\Pi^i}_\alpha e^\alpha$.

The torsion free {H}-valued\footnote{For notational simplicity we do not
distinguish between the group and the algebra here.} connection 
${\omega^\alpha}_\beta$
is then defined by
\begin{equation}de^\alpha +{\omega^\alpha}_\beta \wedge e^\beta=0\end{equation}
and evaluates to
\begin{equation}{\omega^\alpha}_\beta=\left({1\over 2}{f^{\alpha}}_{\beta\gamma} +
{f^{\alpha}}_{\beta i}{\Pi^i}_\gamma\right)e^\gamma.\end{equation}

The curvature 2-forms can then be calculated from
\begin{equation}{R^\alpha}_\beta=d{\omega^\alpha}_\beta + 
{\omega^\alpha}_\gamma\wedge{\omega^\gamma}_\beta
\label{Rab}
\end{equation}
resulting in
\begin{equation}{R^\alpha}_\beta ={1\over 4} 
\left(2{f^\alpha}_{\beta i}{f^i}_{\gamma\delta}
+{f^\alpha}_{\beta\epsilon}{f^\epsilon}_{\gamma\delta}
-{f^\alpha}_{\gamma\epsilon}{f^\epsilon}_{\beta\delta}
\right)e^\gamma\wedge e^\delta.
\label{fab}
\end{equation}
On a symmetric space these reduce to the simpler form
\begin{equation}{R^\alpha}_\beta ={1\over 2} 
\left({f^\alpha}_{\beta i}{f^i}_{\gamma\delta}
\right)e^\gamma\wedge e^\delta,
\label{curvature}
\end{equation}
so the Riemann tensor has components
\begin{equation}{R^\alpha}_{\beta\gamma\delta} = 
{f^\alpha}_{\beta i}{f^i}_{\gamma\delta}.\end{equation}

On a non-symmetric space there is a second, very useful,
connection that comes from introducing a torsion tensor which is
identified with the non-symmetric structure constants:
\begin{equation}{T^\alpha}_{\beta\gamma}={f^\alpha}_{\beta\gamma}\end{equation}
giving torsion 2-forms
\begin{equation}{T^\alpha}={1\over 2}{f^\alpha}_{\beta\gamma}e^\beta\wedge e^\gamma.\end{equation}
Then the connection with torsion is defined via
\begin{equation}de^\alpha +{{\,\omega}^\alpha}_\beta \wedge e^\beta=T^\alpha\end{equation}
which leads to 
\begin{equation}{\,\omega^\alpha}_\beta={1\over 2}
{f^{\alpha}}_{\beta i}{\Pi^i}_\gamma e^\gamma.
\label{Tomega}
\end{equation}
The resulting  curvature 2-forms are
\begin{equation}
{{\, R^\alpha}}_\beta ={1\over 2} 
{f^\alpha}_{\beta i}{f^i}_{\gamma\delta} e^\gamma\wedge e^\delta,
\label{fabT}
\end{equation}
giving curvature tensor 
\begin{equation}{\, R^\alpha}_{\beta\gamma\delta} =
{f^\alpha}_{\beta i}{f^i}_{\gamma\delta}.
\label{RabT}
\end{equation}
The Ricci scalar for the connection with torsion
is then easily evaluated as
\begin{equation}
R={R^{\alpha\beta}}_{\alpha\beta}={f^{\alpha\beta}}_{i}{f^i}_{\alpha\beta}
={f^{AB}}_{i}{f^i}_{AB}-{f^{jk}}_{i}{f^i}_{jk},
\label{Ricci}
\end{equation}
which can be determined using the appropriate quadratic Casimirs of $H$.

A particular instance of this is when $H$ is trivial so $G/H\cong G$.
Then ${\,\omega^\alpha}_\beta=0$ and 
${\, R^\alpha}_\beta =0$, all co-variant
derivatives are trivial and $T^\alpha$ is called the 
parallelising torsion for $G$.
On a symmetric space, of course, 
(\ref{fab}) and (\ref{fabT}) are identical
because ${f^\alpha}_{\beta\gamma}=0$.

In fact it is not difficult to show, using (\ref{Tomega}), (\ref{fabT})
and the Jacobi identity, that 
${\, R^\alpha}_{\beta\gamma\delta}$ in (\ref{RabT}) is co-variantly
constant,
\begin{equation}\nabla_\epsilon {\, R^\alpha}_{\beta\gamma\delta}=0.\end{equation}

On a generic $d$-dimensional manifold the curvature
2-forms (\ref{Rab}) are $SO(d)$
Lie algebra valued 2-forms, but on $G/H$ both (\ref{fab}) and (\ref{fabT})
are {H} valued
2-forms, where $H\subseteq SO(d)$.  This means that we can take
linear combinations of (\ref{fabT}) 
that lie in {H} without
losing any information.
For example, if $H$ is semi-simple, taking the combination
\begin{equation}
{f^i}_{\alpha\beta}R^{\alpha\beta}={1\over 2}
\Bigl( C_2(G,adj) - C_2(H,adj)\Bigr)
{f^i}_{\gamma\delta}\e^\gamma\wedge\e^\delta 
\end{equation}
suggests defining
\begin{equation}
F^i:=
{1\over 2}{f^i}_{\gamma\delta} e^\gamma\wedge e^\delta
\end{equation}
and then $F^i$ are {H}-valued 2-forms which are equivalent to
(\ref{fabT}) (this formula is easily adapted to the case where
$H$ contains $U(1)$ factors).

\section{Appendix}

The calculation of the full spectrum of the Dirac operator
on $CP^2$ proceeds as follows.
For $SU(3)$ the second order Casimir and dimension are
\begin{equation}
C_2(p,\bar p)={1\over 3}\Bigr(p(p+3) + \bar p(\bar p+3) + p\bar p\Bigl)
\end{equation}
and
\begin{equation}
d(p,\bar p)={1\over 2}(p+\bar p +2)(p+1)(\bar p+1)
\end{equation}
respectively.
With $p=r+s$ and $\bar p=\bar r+\bar s$, as in the text,
the constraints read
\begin{equation}
(s -\bar s) -2(r-\bar r)=Y \qquad \hbox{and} \qquad{s+\bar s\over 2}=I,
\end{equation}
where $Y$ is even (odd) for $I$ integral (half-integral).
Now the spectrum depends on whether $|Y|\ge 2I$ or  $|Y|\le 2I$:

\bigskip 
\noindent $\bullet$ If $Y\ge 2I$ then $\bar r \ge r$:
in this case let $n= \bar s$, so $n=0,\ldots,2I$, $r=k$ and 
$\bar r=k+n-I+{Y\over 2}$,
for $k$ a non-negative integer.
If $Y\le -2I$ then $r\ge \bar r$:  in this case let
$n= s$, so $n=0,\ldots,2I$,
$r=k+n-I-{Y\over 2}$ and $\bar r=k$, for $k$ a non-negative integer.
In either case
\begin{equation}
C_2(p,q)=k\left(k+n+2+I+{|Y|\over 2}\right) 
+n\left(n+1-I+{|Y|\over 2}\right)+{|Y|\over 2}+{Y^2\over 12}
+I(I+1).
\end{equation}
For $C_2(H,t_i)$ in (\ref{Eigenvalues}) take the $U(1)$ background
to have fixed charge $M$ 
and the $SU(2)$ background to have
isospin $J$, so 
\begin{equation}
C_2(H,t_i)={M^2\over 12}+J(J+1),
\end{equation}
(the ${1\over 12}$ here is because the $U(1)$ gauge field is
a multiple of ${1\over 2\sqrt 3}$ to conform with the normalisation
of $t_8$ in appendix C).
Finally the Ricci scalar for $CP^2$ can be evaluated from
(\ref{Ricci}) and
the structure constants in appendix C 
to be $R=6$ so, putting all this together,
the eigenvalues of (\ref{Eigenvalues}) are
\begin{eqnarray}
E(k,n)&=&k\left(k+n+2+I+{|Y|\over 2}\right) 
+n\left(n+1-I+{|Y|\over 2}\right)\nonumber\\
&&\qquad\quad+{(|Y|+3)^2\over 12} -{M^2\over 12}
+I(I+1)-J(J+1), 
\end{eqnarray}
while the degeneracies are
\begin{equation}
d(k,n)={1\over 2}\left(2k+n+I+2+{|Y|\over 2} \right)
\left(k+2n-I+1+{|Y|\over 2} \right)\left(k-n+2I+1 \right),
\label{Yge2I}
\end{equation}
with $n=0,\ldots,2I$ and $k\ge 0$ an integer.

It is important to understand how the gauge charges $M$ and $J$ 
are related to the total charges 
$Y$ and $I$ (which include the spin connection).
There are four cases to consider:
\begin{enumerate}
\item  $M=Y\pm 3$ and $I=J$, these are states that couple to
the $U(1)$ part of the spin connection and not the $SU(2)$ part (the
$\pm 3$ relates to the fact that the first Chern class of the tangent
bundle for $CP^2$ is 3).  The spectrum is
\begin{equation}
E(k,n)= k\left(k+n+2+I+{|Y|\over 2}\right) 
+n\left(n+1-I+{|Y|\over 2}\right)+{|Y|\mp Y\over 2};\nonumber \\
\label{spindown}
\end{equation}
For $SU(2)$ singlets $I=J=0$, so $n=0$, this spectrum agrees with the
results of \cite{GrosseStrohm} (in the
notation of that reference $M=2m+3$, so $Y/2=m$ or $m+3$).
 
\item  $M=Y$ and $I=J\pm {1\over 2}$, these are states that couple to
the $SU(2)$ part of the spin connection and not the $U(1)$
part.  The spectrum is  
\begin{eqnarray}
E(k,n)&=& k\left(k+n+2+I+{|Y|\over 2}\right) 
+n\left(n+1-I+{|Y|\over 2}\right)+{|Y|\over 2}+I+1;\nonumber\\
&&\label{spinup1}\\ 
E(k,n)&=& k\left(k+n+2+I+{|Y|\over 2}\right) 
+n\left(n+1-I+{|Y|\over 2}\right)+{|Y|\over 2}-I.\nonumber\\
\label{spinup2}
\end{eqnarray}
\end{enumerate}
For $|Y|>2I$ only case 1 above allows for zero-modes (when $k=n=0$).
For $|Y|=2I$ there are zero-modes in both cases.

\bigskip
\noindent $\bullet$ If $0\le Y\le 2I$, let $n={\bar s-s +Y\over 2}$,
so $n=-I+{Y/2},\ldots,I+{Y/2}$.  Then: either $r=k+|n|$ and $\bar r=k$;
or $r=k$ and $\bar r= k+|n|$. 
If $-2I\le Y\le 0$, let $n={s-\bar s -Y\over 2}$,
so $n=-I-{Y/2},\ldots,I-{Y/2}$.  Then: either $r=k+|n|$ and $\bar r=k$;
or $r=k$ and $\bar r= k+|n|$.
In either case:
\begin{equation}
C_2(p,q)=k\left(k+2I+2\right) 
+n^2+|n|(k+I+1)-{n\over 2}|Y|+{Y^2\over 12}
+I^2+2I.
\end{equation}
The eigenvalues of (\ref{Eigenvalues}) are therefore
\begin{eqnarray}
E(k,n)&=&k\left(k+2I+2\right) 
+n^2+|n|(k+I+1)-{n\over 2}|Y|\nonumber\\
&&\qquad\qquad +{Y^2-M^2\over 12}+I^2+2I-J(J+1)+{3\over 4},
\end{eqnarray}
with degeneracies
\begin{eqnarray}
d(k,n)&=&\nonumber\\
&&\kern -55pt \left\{(k+I+1)^2 +|n|(k+I+1) -{(4n-|Y|)(2n-|Y|)\over 4} \right\}
\left(k+I+1+{|n|\over 2}\right),\nonumber\\
\label{Yle2I}
\end{eqnarray}
where $-I+{|Y|\over 2}\le n\le I+{|Y|\over 2}$ and $k\ge 0$.
(The degeneracy is always an integer because of the restriction that
$Y$ is odd when $I$ is half-integral and even if $I$ is integral.)

Again there are four possibilities:
\begin{enumerate}
\item  $M=Y\pm 3$ and $I=J$, 
\begin{equation}
E(k,n)=k\left(k+2I+2\right) 
+n^2+|n|(k+I+1)-{n|Y|\over 2}+I\mp{Y \over 2};\\
\end{equation}
\item  $M=Y$ and $I=J\pm {1\over 2}$, 
\begin{eqnarray}
E(k,n)&=&k\left(k+2I+2\right) 
+n^2+|n|(k+I+1)-{n\over 2}|Y|+2I+1;\nonumber\\
&&\label{spinup3}\\
E(k,n)&=&k\left(k+2I+2\right) 
+n^2+|n|(k+I+1)-{n\over 2}|Y|.
\label{spinup4}
\end{eqnarray}
\end{enumerate}
These eigenvalues are bounded below by zero and, for $|Y|<2I$, 
only (\ref{spinup4}) allows for zero-modes (when $k=n=0$).
When $|Y|=2I$ the spectrum agrees with equations 
(\ref{spindown})-(\ref{spinup2}). 

\section{Appendix}

In this section we give an explicit  evaluation of the index
of the Dirac operator on $SU(3)/U(1)\times U(1)$, using
standard differential-geometric techniques. 

Let $\lambda_A$; $A=1,\ldots ,8$ be the Gell-Mann matrices
for $SU(3)$, so
\begin{equation}
[t_A,t_B]=i{f_{AB}}^C t_C \qquad\hbox{with}\qquad t_A={\lambda_A\over 2}
\end{equation}
and
\begin{equation}
f_{123}=1, \quad 
f_{453}=-f_{673}=f_{471}=-f_{561}=f_{462}=f_{572}={1\over 2}, \quad
f_{458}=f_{678}={\sqrt 3\over 2}.
\end{equation}
In the notation of appendix A, $i=3,8$ and $\alpha=1,2,4,5,6,7$,
when 
\begin{equation}
t_3={1\over 2}\left( \matrix{1&0&0\cr 0&-1&0\cr 0&0&0\cr }\right)
\qquad\hbox{and}\qquad
t_8={1\over 2\sqrt 3}\left( \matrix{1&0&0\cr 0&1&0\cr 0&0&-2\cr }\right)
\end{equation}
are chosen as the $U(1)\times U(1)$ generators.  This space is
not symmetric, because some of the $f_{\alpha\beta\gamma}\ne 0$.
The curvature 2-forms, for the spin connection with torsion
described in appendix A, are 
\begin{eqnarray*}
R_{12}&=&{1\over 2}(2\,\e^1\wedge\e^2 + \e^4\wedge\e^5 -\e^6\wedge\e^7)
\nonumber\\
R_{45}&=&{1\over 2}(\e^1\wedge\e^2 + 2\,\e^4\wedge\e^5 +\e^6\wedge\e^7)
\nonumber\\
R_{67}&=&{1\over 2}(-\e^1\wedge\e^2 + \e^4\wedge\e^5 +2\,\e^6\wedge\e^7).
\nonumber\\
\end{eqnarray*}
These are not independent, since $R_{45} = R_{12} + R_{67}$,
they are associated with two $U(1)$ field strengths:
\begin{eqnarray*}
F^{3}t_3&=&{1\over 2}{f_{\alpha\beta}}^3\bigl(\e^\alpha\wedge\e^\beta\bigr) t_3=
{1\over 4}(2\,\e^1\wedge\e^2 + \e^4\wedge\e^5 -\e^6\wedge\e^7)
\left(\matrix{1&0&0\cr 0&-1&0\cr 0&0&0\cr}\right)
\nonumber\\
F^{8}t_8&=&{1\over 2}{f_{\alpha\beta}}^8\left(\e^\alpha\wedge\e^\beta\right) t_8=
{1\over 4}(\e^4\wedge\e^5 +\e^6\wedge\e^7)
\left(\matrix{1&0&0\cr 0&1&0\cr 0&0&-2\cr}\right).
\nonumber\\
\end{eqnarray*}
We extract $U(1)$ singlets by projecting out the top left-hand components
\begin{eqnarray*}
F_{(3)}&=
&
{1\over 4}(2\,\e^1\wedge\e^2 + \e^4\wedge\e^5 -\e^6\wedge\e^7)
\nonumber\\
F_{(8)}&=&
{1\over 4}(\e^4\wedge\e^5 +\e^6\wedge\e^7).
\end{eqnarray*}
A monopole field with charges $(M,N)$ is a linear combination of
these
\begin{equation}
F=M F_{(3)} + N F_{(8)}
\end{equation}
from which 
\begin{equation}
F\wedge F\wedge F = {3\over 16} M(N^2-M^2)\,
\e^{124567}
\end{equation}
(we use the shorthand $\e^{124567}=\e^1\wedge\e^2\wedge\e^4\wedge\e^5\wedge\e^6\wedge\e^7$).
The index of the Dirac operator is \cite{AGW}
\begin{equation}
\nu={1\over (2\pi)^3} \int \left\{ {1\over 6}F\wedge F\wedge F
+{1\over 48} F\wedge tr(R\wedge R)\right\}.
\end{equation}
Explicit calculation reveals that $F\wedge tr(R\wedge R)=0$, so
\begin{eqnarray*}
\nu={1\over 256 \pi^3} M(N^2 - M^2) {\mathit V},
\end{eqnarray*}
where ${\mathit V}=\int  \e^{124567}$
is the volume of $SU(3)/U(1)\times U(1)$.
The normalisation can be fixed by using the fact 
$SU(3)/ U(1)\times U(1)$ has Euler characteristic
$\chi=6$, so
\begin{equation}
\chi={1\over 3!}{1\over (4\pi)^3} 
\int \epsilon^{\alpha_1\cdots\alpha_6} R_{\alpha_1\alpha_2}\wedge
R_{\alpha_3\alpha_4}\wedge R_{\alpha_5\alpha_6}=6.
\end{equation}
this fixes ${\mathit V}=32\pi^2$ so
\begin{equation}
\nu ={1\over 8}M(N^2-M^2).
\label{su3index}
\end{equation}
Note that $M$ and $N$ must be either both even or both odd for
$\nu$ to be an integer.


\begin{thebibliography}{03}

\bibitem{Girvin} S.M.~Girvin, ``The Quantum Hall Effect: Novel Excitations 
and Broken Symmetries'', Les Houches lectures (1998), (cond-mat/9907002)

\bibitem{Bernevig} B.A,~Bernevig, J.~Brodie, L.~Susskind and N.~Toumbas,
{\it JHEP} {\bf 0102} (2001) 003, (hep-th/0010105)

\bibitem{ZhangHu} Shou-Cheng Zhang and Jiangping Hu,
{\it Science} {\bf 294} (2001) 823,
(cond-mat/0110572); Jiangping Hu and Shou-Cheng Zhang,
``Collective Excitations at the Boundary of a 4D
Quantum Hall Droplet'', (cond-mat/0112432)

\bibitem{Haldane} F.D.M.~Haldane, {\it Phys.~Rev.~Lett.} {\bf 51} (1983)
605

\bibitem{HouPeng} Yi-Xin Chen, Bo-You Hou and Bo-Yuan Hu,
``Non-commutative Geometry of 4-dimensional Quantum Hall Droplet'', 
(hep-th/0203095);
B.A.~Bernevig, Chyh-Hong Chern, Jian-Ping Hu, N.~Toumbas and Shou-Chen Zhang,
``Effective field Theory Description of the Higher Dimensional Quantum Hall
Liquid'', (cond-mat/0206164);
H.~Elvang and J.~Polchinski, ``The Quantum Hall
Effect on $R^4$, (hep-th/0209104);
Bo-You Hou and Dan-Tao Peng, ``Incompressible Quantum Hall Fluid'', 
(hep-th/0210173)

\bibitem{KarabaliNair} D.~Karabali and V.P.~Nair, {\it Nuc.~Phys.}
{\bf B641}~(2002)~533, (hep-th/0203264)

\bibitem{Fabinger} M.~Fabinger, ``Higher-dimensional Quantum Hall
Effect and String Theory'', (hep-th/0201016)

\bibitem{susyqm} F.~Cooper, A.~Khare and U.~Sukhatme, {\it 
Phys.~Rep.} {\bf 251} (1995), (hep-th/9405029)

\bibitem{WA} M.~Atiyah and E.~Witten, 
{\it Adv.~Theor.~Math.~Phys.} {\bf 6} (2003) 1, (hep-th/0107177)


\bibitem{BGJP} A.P.~Balachnadran, G.~Immirzi, J.~Lee and P.~Pre\u{s}najder, \lq\lq Dirac Operators on Coset Spaces'',
(hep-th/0210297)

\bibitem{EGH} T.~Eguchi, P.~Gilkey and A.~Hanson, {\it Phys.~Rep.} {\bf 66C} (1980) 213

\bibitem{Yang} Chen~Ning~Yang, {J.~Math.~Phys.} {\bf 19} (1978) 320;
{\it ibid.} {\bf 19} (1978) 2622 

\bibitem{BPST} A.A.~Belavin, A.M.~Polyakov, A.S.~Schwartz and Y.S.~Tyupkin,
{\it Phys.~Lett.} {\bf 59B} (1975) 85

\bibitem{HP} S.~Hawking and C.~Pope, {\it Phys.~Lett.} {\bf 73B} (1978) 42

\bibitem{DolanNash} B.P.~Dolan and C.~Nash, 
{\it JHEP} {\bf 0207} (2002) 057, (hep-th/0207007)

\bibitem{KobayashiNomizu} Shoshichi Kobayashi and
Katsumi Nomizu, {\it Foundations of Differential Geometry}, (1963), (Wiley)

\bibitem{SalamStrathdee} A.~Salam and Strathdee, 
{\it Ann.~of Phys.} {\bf 141} (1982) 316
\bibitem{AGW} L.~Alvarez-Gaum\'e and P.~Ginsparg, {\it Ann.~Phys.} {\bf 161}
(1985) 423

\bibitem{GrosseStrohm} H.~Grosse and A.~Strohmaier,  
{\it Lett.~Math.~Phys.} {\bf 48} (1999) 163 (hep-th/9902138)

\end{thebibliography}
\end{document}